# NEUTRON STARS, THE MOST EXOTIC NUCLEAR LAB IN THE UNIVERSE

PIERRE M. PIZZOCHERO

*Dipartimento di Fisica, Università degli Studi di Milano, and
Istituto Nazionale di Fisica Nucleare, sezione di Milano,
Via Celoria 16, 20133 Milano, Italy*

**ABSTRACT**

In this lecture, we give a first introduction to neutron stars, based on fundamental physical principles. After outlining their amazing macroscopic properties, as obtained from observations, we infer the extreme conditions of matter in their interiors. We then describe two crucial physical phenomena which characterize compact stars, gravitational stability of strongly degenerate matter and neutronization of nuclear matter with increasing density, and explain how the formation and properties of neutron stars are a consequence of the extreme compression of matter under gravity. Finally, we describe how astronomical observations of various external macroscopic features can give invaluable information about the exotic microscopic scenario inside: neutrons stars represent a unique probe to study super-dense, isospin-asymmetric, superfluid, bulk hadronic matter.

## I – INTRODUCTION

Neutron stars are among the most exotic objects in the Universe: indeed, they present extreme and quite unique properties both in their macrophysics, controlled by the long-range gravitational and electromagnetic interactions, and in their microphysics, controlled by the short-range weak and strong nuclear interactions. Actually, neutron stars are as astounding to the astrophysicist as they are to the nuclear physicist: what the former sees as an incredibly compact star appears to the latter as an incredibly extended nucleus. Their extreme stellar properties (e.g., gravitational potential, rotational frequency, magnetic field, surface temperature) are matched by their extreme nuclear properties (e.g., nucleon density, isospin asymmetry, bulk superfluidity, neutrino reactions).

A comprehensive study of neutron stars requires an understanding of how the long and short range interactions affect each other. Indeed, the physics of compact stars and of the stellar systems they form constitutes a new and thriving field of research, *relativistic nuclear astrophysics* (also called *astronuclear* physics), which requires expertise from disciplines that are generally mostly disjointed, but now have to work sinergically: high-energy astrophysics, gravitational physics (i.e., general relativity), nuclear and hadronic physics, neutrino physics, QCD, physics of superfluids. Therefore, presenting the subject of neutron stars to students is not a simple task: the interdisciplinary nature of this study makes it a long, although rewarding, process. Several books and many reviews on the different topics involved are presently available [1], and regular schools are organized every year. In Europe, the ESF network *Compstar* [2] is dedicated to the physics of compact stars and to the formation of young astronuclear physicists.

Obviously, the goal of this lecture is not to present a complete overview of neutron stars, of their properties and of the physical processes involved: on top of being impossible in such a short lecture format, this would be useless to an audience of nuclear physicists, who are trained in the physics of microscopic systems of degenerate nucleons, but are probably not familiar with astrophysical issues in general, and with the macroscopic behaviour of objects strongly bound by gravity in particular. Rather, here we introduce the subject of neutron stars to such an audience by first showing how some of their observed basic properties, and the internal physical conditions we infer from them, are easily explained as a consequence of the strong gravitational field which, in this final stage of stellar evolution, compresses matter to its very limit; actually, it is the dominant effect of gravity on the other interactions which characterizes the physics of compact stars and the stellar systems they form. We then explain why observations of neutron stars, and thence information about the exotic environment they present in their interiors, can be invaluable to nuclear physics, since they provide a unique astrophysical laboratory to test the properties of hadronic matter under extreme conditions, not attainable in any terrestrial facility now or in the future.

The lecture develops as a series of answers to four basic questions. In Section II, we introduce the main observed properties of neutron stars and deduce from them their internal physical conditions. In Section III, we briefly describe the gravitational stability of degenerate fermions and the neutronization of matter during gravitational contraction, with which we can explain the origin of neutron stars, their mass range and the conclusion that these stars are basically macroscopic assemblies of neutrons bound by gravity. In Section IV, we outline the composition of hadronic matter along the neutron star profile, from the superfluid crust to the super-dense exotic core, as determined by the increasing density with depth. In Section V, we choose as examples three significant classes of observations (maximum mass, surface cooling and pulsar glitches), and we outline how these observations can yield invaluable information about some fundamental nuclear physics issues, in the present case: the equation of state (EOS) of dense bulk matter, macroscopic nucleon superfluidity, and the properties of neutron-rich exotic nuclei above neutron-drip.

## II – WHAT ARE NEUTRON STARS?

Although predicted theoretically soon after the discovery of the neutron in 1932, neutron stars were first observed accidentally in 1967 in the radio band, as periodically pulsating sources called *pulsars*; since then, about two thousands have been discovered. These radio-pulsars have periods in the range $P \sim 10^{-3} - 10\, s$, and although they gradually slow down with period derivatives in the range $\dot{P} \sim 10^{-20} - 10^{-10}\, s\, s^{-1}$ (see Figure 1a), their periods are so stable that pulsars can be considered the most accurate clocks in the Universe. Coupled to independent estimates of the mass range (see later), a simple analysis of the pulsed emission implies that pulsars must be *rotating compact objects* with radii of the order of $R = 10^6\, cm$. Indeed, among periodic mechanisms, stellar pulsations cannot explain at the same time the slow and fast pulsars, while binary systems losing energy would spin up instead of slowing down. This leaves only rotations, but the shortest periods require very small objects: a white dwarf (typical radius $\sim 10^4\, km$) with a period of one millisecond would be above the keplerian limit for the maximum angular frequency $\omega_k = 2\pi/P_k = \sqrt{GM/R^3}$ (namely, gravitational field strong enough to provide the centripetal acceleration) and would fly apart; a black hole, instead, has no surface and thence cannot emit periodically. The identification with *neutron* stars of these compact objects, of $\sim 1\, M_\odot$ mass (the solar mass is $M_\odot = 2 \times 10^{33}\, g$) and $\sim 10\, km$ radius, will be explained in Section III.

The emission from radio-pulsars was soon interpreted in terms of rotating neutron stars with large dipolar magnetic fields: the periodic emission would then be explained as the beamed radiation emitted by a rotating (i.e., accelerated) dipole. This is the so-called "lighthouse model" (see Figure 1b), since the rotation of the quite narrow radiation beam is seen by the distant observer (if in the right line of sight) as a pulsation; it was later supported by the evidence of cyclotron lines in the spectra, corresponding to huge magnetic fields of order $B \sim 10^{12}\ G$. The dipole model determines the age, $\tau$, and magnetic field of a pulsar from the measured values of $P$ and $\dot{P}$; for example, $\tau = P/2\dot{P}$ (lines of constant $\tau$ and $B$ are shown in Figure 1b). Although model dependent, observations yield the ranges $\tau \sim 10^3 - 10^{10}\ yr$ and $B \sim 10^7 - 10^{16}\ G$, with some of the youngest objects having the largest magnetic fields (*magnetars*) and the oldest ones having the fastest rotation (*millisecond pulsars*); the large majority of pulsars, however, can be found in the range $\tau \sim 10^5 - 10^9\ yr$ and $B \sim 10^{11} - 10^{13}\ G$. These different classes of stars are generally explained in terms of stellar evolution; for example, millisecond pulsars, which are mostly found in binary systems, are interpreted as old neutron stars with a decayed magnetic field that have been spun up by accretion of mass from the companion star.

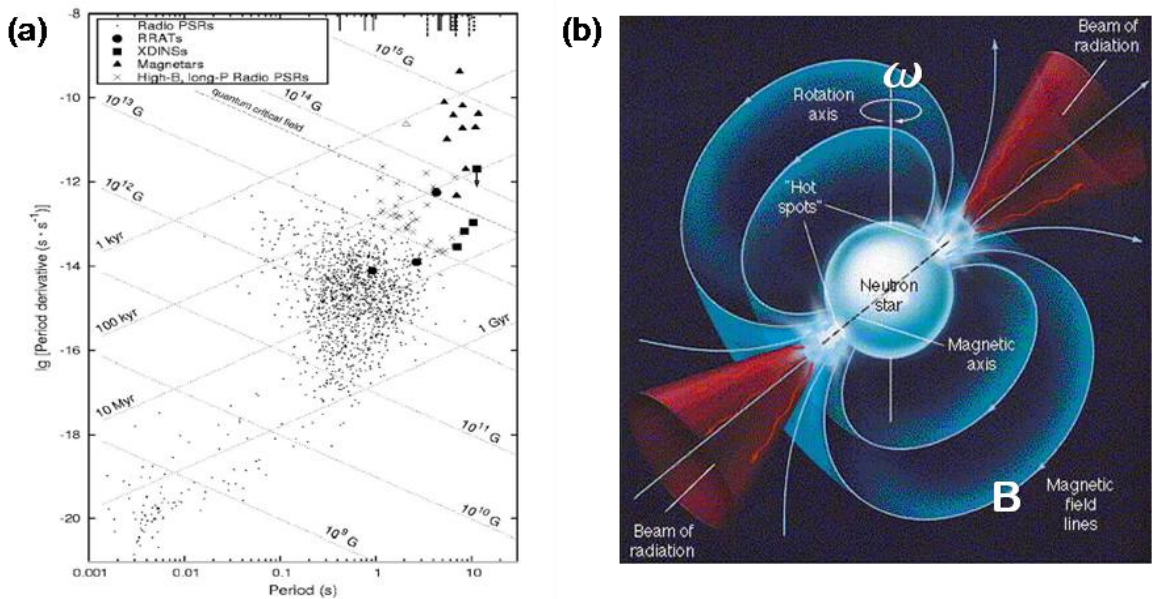

*Figure 1 – (a) $P$-$\dot{P}$ diagram for observed neutron stars. Lines of constant age and constant magnetic field are drawn, according to the rotating dipole model. (b) The lighthouse model for pulsar periodic emission from a rotating magnetic dipole.*

Neutron star masses can be measured when they belong to binary systems, using Kepler's third law coupled to other orbital data [3]. In particular, double neutron star binaries present observable general relativistic effects, due to the very strong gravitational field, that allow a very precise determination of the mass. Figure 2a reveals a mass range $M \sim 1 - 2\ M_\odot$, with an error-weighted mean value (dashed vertical line) of $M \simeq 1.4\ M_\odot$. The best determined masses lie in a narrow interval $M \sim 1.25 - 1.45\ M_\odot$, while values $M \gtrsim 2\ M_\odot$ are affected by large error bars.

Neutron star radii can only be determined indirectly [3]. For example, if the distance of the star from Earth is known, its radius can be obtained by fitting model atmospheres to the observed spectra; or, if the mass is known, it can be inferred by the gravitational redshift of spectral lines. Although model-dependent, the estimated radii lie in a range $R \sim 10 - 15\ km$, consistent with the keplerian limit discussed earlier.

Surface temperatures can be measured for several isolated neutron stars of different ages (see Figure 2b), whose thermal emission is not hidden by strong magnetospheric emission, like it is usually the case [3]. We point out the large error bars in the results: indeed, fitting the observed spectra (e.g., with a black-body or with a model atmosphere) and extracting an effective surface temperature or luminosity is not a model-independent procedure; similarly, the age of the pulsar is determined from the rotating dipole model, which is obviously only a first approximation to the complex and still poorly understood pulsar emission mechanism, actually associated to the extended magnetosphere (region where the dynamics is dominated by the magnetic field) of charged particles strongly coupled to the magnetic field and co-rotating with the star. In spite of the uncertainties, neutron stars appear to cool down as they age, their surface temperatures spanning the interval $T_s \sim 4 \times 10^5 - 2 \times 10^6\,K$, namely they shine in the soft X-ray range.

Incidentally, these temperature observations can only be made from telescopes mounted on satellites, since the Earth's atmosphere is opaque to most electromagnetic waves outside the visible range, except for a large transparency window in the radio frequency range, a fact which made the early discovery of radio-pulsars possible. Nowadays, pulsars are observed in all the different bands of the electromagnetic spectrum and their pulsed emission can also be detected at the IR, visible, UV, X and $\gamma$ energies, in several possible combinations: the Crab and Vela pulsars, for example, pulse in all bands, Geminga is a $\gamma$-ray pulsar, the XDINs (X-ray-dim isolated neutron stars) are radio-quiet and so on.

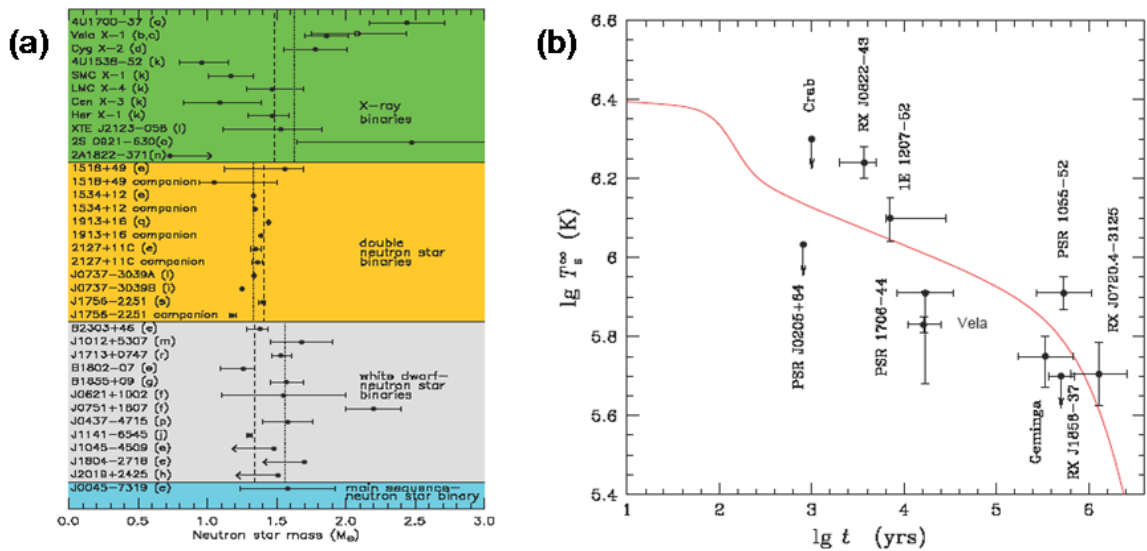

*Figure 2 – (a) Measured masses of neutron stars in binary systems; the data are divided according to the nature of the companion star in the system (from Ref. [3-b]). (b) Measured surface temperatures of isolated neutron stars; the ages reported in abscissa are obtained from the rotating dipole model, and the solid line represents a standard cooling simulation (from Ref. [6-b]).*

We thus see how observations of several properties of pulsars point to quite extraordinary objects from the astrophysical point of view: stars with a mass similar to that of the Sun but constrained by gravity in a sphere with diameter of about 20-30 kilometers, spinning around their axis up to a thousand times per second, shining in the soft X-ray, with a powerful magnetosphere generated by magnetic fields more than a billion times larger than that of the Sun. This magnetosphere, which is like an extension of the neutron star itself, being dragged along by the rapidly rotating star and magnetically containing intense flows of relativistic charged particles, is responsible for both the remarkable pulsed emission and the much larger loss of energy by magnetic dipole radiation and particle emission (pulsar wind). For the rotationally-powered pulsars we are considering here, the bulk loss of energy is made evident by the star's slow-down. In the case of the Crab pulsar, the evidence is even more direct: found in the middle of the Crab Nebula, which is the remnant of the supernova exploded in 1054 and observed by Chinese astronomers, the Crab pulsar powers the Nebula which surrounds it and maintains it very luminous, even a thousand years after the supernova explosion. Indeed, the decrease in rotational kinetic energy of the Crab pulsar, $\dot{K} = I\omega\dot{\omega}$ (within a factor of order unity, the momentum of inertia is $I \simeq MR^2 \sim 10^{45}\,g\,cm^2$, when evaluated for typical neutron star parameters), matches the luminosity of the Crab Nebula ($L_{\text{Crab}} \sim 5 \times 10^{38}\,erg\,s^{-1}$).

To understand why neutron stars are extraordinary also from the microscopic point of view, we must try to infer their unseen internal properties from the observed astrophysical parameters, and thence determine the physical state of matter in their interiors. The average mass density corresponding to $M = 1.4\ M_\odot$ and $R = 15\ km$ is $\bar\rho = 3M/4\pi R^3 \simeq 2 \times 10^{14}\ g\ cm^{-3}$; this is almost the value of nuclear saturation density ($\rho_0 = 2.8 \times 10^{14}\ g\ cm^{-3}$), while stars with $R = 10\ km$ would have $\bar\rho \simeq 2.3\ \rho_0$. Due to the steep gradients in pressure and density necessary to stabilize the star against gravity, we expect much larger central densities; for example, the simple polytropic model, a first approximation to represent gravitationally bound spheres of degenerate fermions, predicts $\rho_c = 6\bar\rho$ for non-relativistic particles and $\rho_c = 54\bar\rho$ for ultra-relativistic ones. We are thus dealing with super-dense matter, compressed by gravity well beyond the average density of normal nuclei: nowhere else in the present Universe can matter with such properties be found and observed. We have the unique scenario of extended (bulk) compressed nuclear matter; in Section III we will argue that it is also strongly isospin-asymmetric, being mostly composed of neutrons (less than one proton every hundred neutrons). Since a solar mass of matter contains $N_\odot = M_\odot/m_N \simeq 10^{57}$ nucleons (where $m_N$ is the nucleon mass), neutrons stars could also be viewed as giant kilometer-sized nuclei with $A \simeq 10^{57}$ and $Y_e = Z/A \lesssim 10^{-2}$ (in astrophysics, the ratio $Z/A$ is indicated by $Y_e$ since it also represents the number fraction of electrons with respect to nucleons); such a point of view is somewhat misleading, however, since these objects are *held together* by gravity alone and *not* by nuclear forces (which, incidentally, are not even able to bind two neutrons together).

To further study the state of matter inside neutron stars, let us take a central density $\rho = 2\rho_0$, but remember that this is conservative and that in the central regions conditions are more extreme. The number density of nucleons is $n = \rho/m_N = 2n_0 \simeq 0.34\ fm^{-3}$ (where $n_0 = \rho_0/m_N = 0.17\ fm^{-3}$ is the nucleon density at saturation) and thence their average distance is $\bar l \sim n^{-1/3} \sim 1.4\ fm$; on the other hand, the thermal De Broglie wavelength of the nucleons is $\lambda_T \sim h/\sqrt{m_N k_B T} \sim 4 \times 10^6\ T^{-1/2}\ fm$. Temperatures inside neutron stars can be inferred indirectly from cooling calculations (see Section V) to reach $T < 10^9\ K$ shortly after the star birth, independent from the initial temperature of formation (which is estimated of order $10^{11}\ K$). This implies that $\lambda_T > 10^2\ fm$ so that $\lambda_T \gg \bar l$. From energy conservation, moreover, an upper limit on the initial temperature can be easily estimated: even if all the gravitational energy released by the collapse of a star to form a neutron star (see Section III) was *integrally* converted into thermal energy of the nucleons, which in reality is obviously not the case, we would have at most $k_B T_{max} \sim GMm_N/R \sim 10^2\ MeV$, corresponding to $T_{max} \sim 10^{12}\ K$; thence the thermal wavelength has the lower limit $\lambda_{T,min} \sim 4\ fm$ which is still larger than $\bar l$.

Therefore, in spite of very high temperatures, up to a billion degrees for very young neutron stars (for a comparison, the interior of the Sun has $T_\odot \sim 10^7\ K$), and due to the extremely large densities, we obtain the strong inequality $\lambda_T \gg \bar l$ in neutron stars interiors; this indicates that a classical description of particles is not valid in such a regime, since their quantum wave-like properties, $\lambda_T$, cannot be neglected at the typical particle scale, $\bar l$. In these conditions, gravity has compressed the nucleons so much that they are even closer than they usually are under the action of nuclear forces alone; they occupy the lowest available energy states and the temperature, although very large, is not anymore a measure of the average particle energy; the degeneracy (zero-point) energy, which is due to the Pauli exclusion principle and is density-dependent, now completely dominates the thermal contribution, $k_B T$. In conclusion, baryonic matter in neutron stars is *strongly* degenerate and a zero-temperature approximation is physically meaningful.

To proceed, we take the Fermi model at $T = 0$ as a guide, namely assume that the nucleons occupy all the lowest momentum states available in phase space up to a maximum value, the Fermi momentum $p_F$. We then have the standard result $p_F = \hbar k_F$, with the Fermi wavenumber $k_F = (3\pi^2 n)^{1/3} \simeq 1.7\ (n/n_0)^{1/3}\ fm^{-1}$. For non-relativistic particles (see later on), we find the Fermi energy $E_F = p_F^2/2m_N \simeq 59\ (n/n_0)^{2/3}\ MeV$; the condition $\lambda_T \gg \bar{l}$ is equivalent to $E_F \gg k_B T$, as expected for strongly degenerate particles. For $n = 2n_0$ we find $p_F \simeq 430\ MeV/c$ and $E_F \simeq 94\ MeV$, while $T < 10^9\ K$ implies $k_B T < 0.1\ MeV$.

The relativity parameter for the nucleons is $\xi = p_F/m_N c \simeq 0.3\ (n/n_0)^{1/3}$; for $n = 2n_0$ we have $\xi \sim 0.4$, and for *any* reasonable density we still find $\xi \lesssim 1$ ; therefore, the nucleons are not ultra-relativistic ($\xi \gg 1$), but they are not completely non-relativistic ($\xi \ll 1$) either. The importance of this fact will be discussed in the next section.

**III – WHY STARS MADE OF NEUTRONS?**

In order to understand why the compact objects that are observed are made of neutrons, to follow in its main steps the evolutionary process that forms neutron stars and to explain the limited range observed for their masses, we must develop two physical ideas, one macroscopic and the other microscopic, but both related to the presence of a strong gravitational field in compact degenerate stars. The first idea is that the degeneracy pressure of fermions is only able to resist gravity up to a *maximum stellar mass*, called Chandrasekhar's mass. If the star's mass is larger than this limit, the star is not gravitationally stable and it must collapse to a different configuration. The second idea is that above density values of order $\rho_\beta \sim 10^7\ g\ cm^{-3}$, increasing density induces electron capture on protons and prevents neutron $\beta$-decay, namely above $\rho_\beta$ we have an increasing degree of *neutronization of matter*.

The existence of maximum allowed masses for different classes of stars is related to a very general property, following from the gravitational virial theorem: a self gravitating object is bound and stable as long as the particles that provide the pressure to resist gravity are non-relativistic; as the particle energies approach relativistic values, because of temperature or density, the system becomes gravitationally unstable to small perturbations and, for ultra-relativistic particles, it is no longer bound together by gravity. This can be applied to different scenarios; for example, classical (extended) stars tend to become hotter, and thus increasingly dominated by radiation (i.e., ultra-relativistic) pressure, the more massive they are; indeed, stars above $50\ M_\odot$ are very rare and those above $100\ M_\odot$ are exceptions. In the present case of degenerate fermions in compact stars, the gravitational instability is also related to the pressure-providing particles (electrons for white dwarfs, nucleons for neutron stars) becoming relativistic. To better explain the origin of the Chandrasekhar's mass limit for compact stars, the following argument, due to Landau, is as simple as it is revealing.

We consider a star of mass $M$ and radius $R$, made of $N = M/m$ completely degenerate ($T = 0$) fermions of mass $m$. The (average) number density of particles is $n \simeq N/R^3$ and their Fermi momentum is $p_F \simeq \hbar\, n^{1/3} \simeq \hbar\, N^{1/3} R^{-1}$. The total gravitational energy of the star is $E_G \simeq -G\, M^2/R$ (we neglect factors of order unity; for a uniform-density sphere we would have a factor of 3/5) and thus the gravitational energy per particle is $U \simeq -Gm^2 N/R$. The (average) kinetic energy of the particles is just their Fermi energy, namely $K \simeq E_F$ (where again we omit factors of order unity, in the present case 3/5 or 3/4 for non-relativistic or ultra-relativistic particles) where $E_F = p_F^2/2m \simeq \hbar^2/m\ N^{2/3} R^{-2}$ for $\xi \ll 1$ and $E_F = p_F c \simeq \hbar c\, N^{1/3} R^{-1}$ for $\xi \gg 1$. The total energy per fermion is then $E = U + K$ and its behaviour with $R$ determines the stability properties of the star. In Figure 3, we plot $E$ as a function of the star's radius. For large values of $R$, the density is low and the particles are non-relativistic ($\xi \ll 1$), so that $E \simeq -Gm^2 N/R + \hbar^2/m\ N^{2/3} R^{-2} \approx -1/R$ as $R \to \infty$. For small values of $R$, the density is high and the particles are ultra-relativistic ($\xi \gg 1$) so that $E \simeq -Gm^2 N/R + \hbar c\, N^{1/3} R^{-1} \approx (-Gm^2 N + \hbar c\, N^{1/3})/R$ as $R \to 0$. Because of the term in parenthesis, there is a critical value of the total particle number $N_{max} \simeq (\hbar c/Gm^2)^{3/2}$ and of the star's mass $M_{max} \simeq m(\hbar c/Gm^2)^{3/2}$, such that (as $R \to 0$) $E \approx +1/R$ if $M < M_{max}$ and $E \approx -1/R$ if $M > M_{max}$ . The asymptotic behavior just described allows to sketch the *shape* of the curves $E = E(R)$, as shown in Figure 3, where the solid line corresponds to masses $M < M_{\text{max}}$ and the dashed line to $M > M_{\text{max}}$.

The minimum in the curve for $M < M_{\text{max}}$ shows that such stars have a stable equilibrium configuration for some value $R_0$ of the radius; conversely, stars with $M > M_{\text{max}}$ cannot be stabilized against gravity by the pressure of degenerate fermions alone and is bound to collapse. Taking $m = m_N$ as it is the case for neutron stars, where nucleons provide both mass and pressure, we find $N_{\text{max}} \simeq 2 \times 10^{57}$ and $M_{\text{max}} \simeq 1.5\, M_\odot$. The order of magnitude of the equilibrium radius for stars with $M \lesssim M_{\text{max}}$ can be estimated by noting that the departure of the solid curve from the $-1/R$ behavior at large $R$ is due to the particles becoming relativistic, so that $R_0$ must be small enough to obtain $\xi \gtrsim 1$. Since $\xi = p_F/m_N c \simeq \hbar/m_N c\ N^{1/3} R^{-1}$, the value of $R_0$ must satisfy $R_0 \lesssim \hbar/m_N c\ N_{\text{max}}^{1/3} \simeq 20\ km$. It is quite amazing how the macroscopic properties of neutron stars appear naturally from the microscopic behavior of their constituent particles, and that their scale is determined by simple combinations of fundamental constants (e.g., $N_{\text{max}} \simeq \alpha_G^{-3/2}$, where $\alpha_G = G m_N^2/\hbar c \simeq 6 \times 10^{-39}$ is the gravitational equivalent of the fine-structure constant $\alpha = e^2/\hbar c \simeq 1/137$ of electromagnetism, and sets the strength of gravity in baryonic matter).

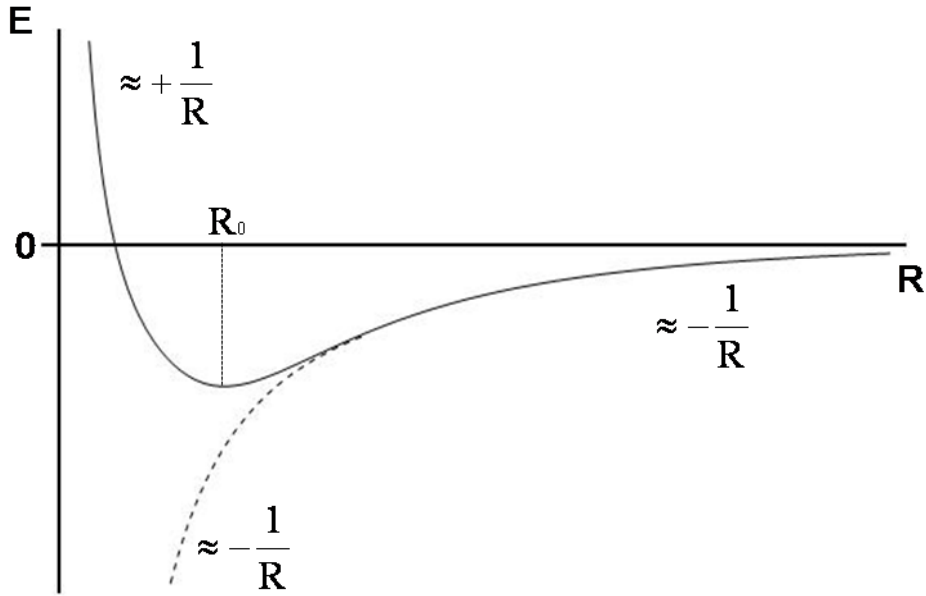

*Figure 3 – Energy per particle in a star of mass M and radius R, whose pressure is provided by N completely ($T = 0$) degenerate fermions. The asymptotic behavior of the energy is indicated for large (non-relativistic regime) and small (ultra-relativistic regime) values of the radius. The solid curve corresponds to $N \lesssim 10^{57}$, and has a stable bound state at $R = R_0$. The dashed line corresponds to $N > 10^{57}$ and leads the star to a complete collapse.*

Of course, the previous argument only sets the scale of the maximum mass of neutron stars, not its precise value; we will come back to this issue in Section V. For now, we just mention the existence of a strong upper limit to the mass of a stable neutron star, $M \lesssim 3\, M_\odot$, derived from General Relativity under very reasonable assumptions (microscopic stability and causality of the EOS of dense matter above saturation).

Landau's argument could be extended to describe *dwarf stars,* namely inert stellar objects stabilized against gravity by the degeneracy pressure of their electrons, and which appear as final stages of stellar evolution (see later on). The argument would yield the same maximum mass, but a planet-like radius of order $\sim 10^4\ km$ and thence average densities of order $\sim 10^6\ g\ cm^{-3}$. At these densities, matter is described by classical (thermal) ions and strongly degenerate electrons, the latter providing most of the pressure. Dwarf stars were first studied theoretically by Chandrasekhar; in the limit $T = 0$ (complete degeneracy), he calculated their limiting mass as $M_{\text{Ch}} = 5.76 \, Y_e^2 \, M_\odot$, where $Y_e = Z/A$ is the electron fraction corresponding to ions of charge $Z$ and mass number $A$. Typically, $Y_e \simeq 1/2$ so that $M_{\text{Ch}} \simeq 1.44 \, M_\odot$; we will come back to this value at the end of this section.

We now turn to discuss the neutronization of matter with increasing density. This is a direct consequence of the shift in $\beta$-equilibrium induced by the presence of increasingly relativistic degenerate electrons, whose total energy grows with density much faster than that of the non-relativistic nucleons. To understand the physical principle, we take the simplest case of capture on protons and study the equilibrium of the weak process $p + e \leftrightarrows n + \nu$ (capture on nuclei will be discussed in Section IV). The thermodynamical equilibrium of protons, neutrons, electrons and neutrinos is expressed by the equality $\mu_p + \mu_e = \mu_n + \mu_\nu$ between their chemical potentials, coupled to the charge-neutrality condition $n_p = n_e$. We now take completely degenerate particles at $T = 0$, as it is the case in compact stars; we then have $\mu_i = E_{F,i} = \sqrt{(m_i c^2)^2 + (p_{F,i} c)^2}$ and $p_{F,i} = \hbar \, (3\pi^2 n_i)^{1/3}$ ($i = n, p, e, \nu$), with $m_i$ the respective rest masses. The complete relativistic expression for the energy *must* be taken, when particle reactions occur and the mass-energy equivalence is crucial in the total energy balance. Since the massless neutrinos escape from the system ($n_\nu = 0$), we take accordingly $\mu_\nu = 0$ and the $\beta$-equilibrium condition becomes $E_p + E_e = E_n$ (where we simplify the notation as $E_i \equiv E_{F,i}$ and $p_i \equiv p_{F,i}$). We stress again how, in this regime, the Fermi energies are density-dependent, but the dependence is different for non-relativistic and ultra-relativistic particles; indeed, we have the standard expansions $E_i \simeq m_i c^2 + p_i^2 / 2m_i$ if $\xi \ll 1$ and $E_i \simeq p_i c$ if $\xi \gg 1$, where $p_i \simeq \pi \hbar \, n_i^{1/3}$.

From this discussion of the n-p-e system, it becomes already clear how neutronization sets in. The mass difference $Q = (m_n - m_p) c^2 \simeq 1.3 \, MeV$ prevents capture of low-energy electrons on protons, since there would be an energy deficit of $\Delta = Q - m_e c^2 \simeq 0.8 \, MeV$. Therefore, at low densities matter in $\beta$-equilibrium is made of protons and electrons only, namely $n_n = 0$: any neutron would have decayed, while no new neutron can be formed by electron capture. However, $\beta$-equilibrium can be shifted to the neutron side whenever the electrons acquire enough kinetic energy to provide the missing $\Delta$. This is possible if matter is dense enough that $E_e > Q$; but a $1.3 \, MeV$ electron is already relativistic, since its kinetic energy ($0.8 \, MeV$) is larger than its rest mass energy ($0.5 \, MeV$). Therefore, we expect neutronization to set in when the density is high enough that electrons become relativistic; under these conditions, not only are electrons captured by protons, but the neutrons thus produced cannot decay back, since the electron levels are already filled up to $E_e$ (Pauli blocking); the neutronized state of matter is therefore *stable*. To find the critical density for the onset of neutronization, we observe that a $1.3 \, MeV$ electron has $p_e c \simeq 1.2 \, MeV$ (so that $\xi_e \simeq 2.4$); this corresponds to $n_e = (p_e / \hbar)^3 / 3\pi^2 \simeq 7 \times 10^{30} \, cm^{-3}$. Since the number density of nucleons is $n = n_p = n_e$, the critical density is $\rho_\beta = n \, m_N \simeq 1.2 \times 10^7 \, g \, cm^{-3}$.

The n-p-e system can be studied exactly; the $\beta$-equilibrium and charge-neutrality conditions can be rewritten as $m_p (1 + \xi_p^2)^{1/2} + m_e (1 + \xi_e^2)^{1/2} = m_n (1 + \xi_n^2)^{1/2}$ and $m_p \xi_p = m_e \xi_e$; the system of equations can then be solved numerically for $n_n / n_p$ as a function of $n = n_p + n_n$. The neutron-to-proton ratio is found to be zero for densities up to $\rho_\beta$. It then increases with increasing density, to reach a maximum value of $n_n / n_p \simeq 385$ at $\rho \simeq 7.8 \times 10^{11} \, g \, cm^{-3}$, and from there it decreases to reach its asymptotic value of $n_n / n_p \to 8$ as $\rho \to \infty$. For $\rho = 2\rho_0$ we find $n_n / n_p \simeq 130$ and $n \simeq n_n$, so that the asymmetry of typical neutron star matter is $Y_e = n_e / n = n_p / n \simeq n_p / n_n \approx 0.008$; calling these compact objects *neutron stars* is indeed quite accurate, with about one proton and one electron every hundred neutrons.

Having shown that the compact objects observed as pulsars have a maximum *allowed* mass of a few solar masses and that they are mostly made of gravity-compressed neutrons, we can now outline how such neutron stars are formed during the last stages of stellar evolution. We will only give the main physical ideas, and refer the interested reader to any standard book on stellar astrophysics [4]. The whole history of a star is based on an ongoing competition between gravity and pressure: as gravity forces matter to contract, different sources of pressure are exploited to counteract the inward pull and stabilize the star until its next evolutionary stage. Therefore, the initial contraction of a large and diffuse cloud of interstellar matter (H, He and traces of heavier elements) down to the formation of a protostar, which itself keeps contracting and heating up, is finally halted by the ignition in the stellar core of a thermonuclear reaction, H burning. The thermal pressure produced by fusion is then able to oppose gravity for a very long time, millions to billions of year, according to the mass (more massive stars have much shorter lives): this phase is called the *main sequence.*

When hydrogen in the core has been transformed into helium, the nuclear reactions stop and the core starts contracting again under the pull of gravity; hydrostatic equilibrium of the system requires the large hydrogen envelope that surrounds the core to expand and thence cool; the process is again halted by the ignition of nuclear reactions, which burn He into C and O. The larger temperatures and densities necessary to fuse heavier nuclei with larger electric charges are thus obtained by gravitational contraction of the stellar core. The star is now a giant object with a colder surface: this is the *red giant* phase. After depletion of He in the core, the process can be repeated over and over: the core is compressed and heated up by subsequent contractions, which are temporarily halted by the fusion of the previous nuclear ashes into heavier elements (the main burning stages being C→Ne→O→Si→Fe). The extent of nuclear burning depends on the initial mass of the star, which actually determines its whole evolutionary history. This is illustrated in Figure 4, where the main different stages of stellar evolution are shown as a function of the mass of the protostar.

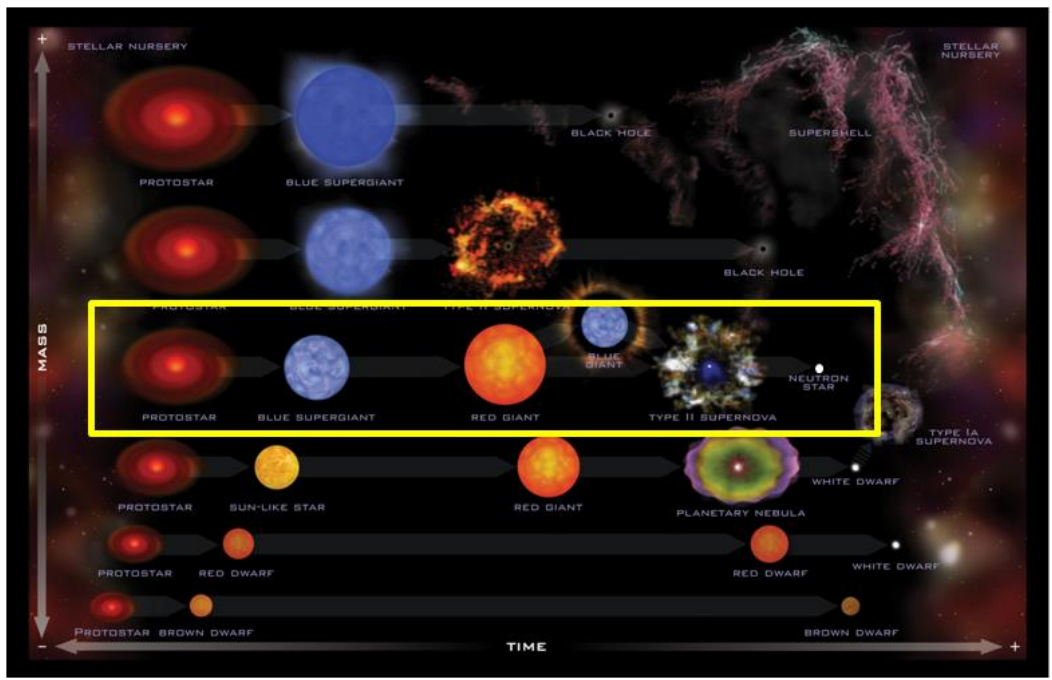

*Figure 4 – Evolutionary stages of stars as a function of the initial mass of the protostar. We highlighted the mass range that leads to neutron stars, namely $8 \lesssim M \lesssim 25\ M_\odot$, with the following stellar stages: protostar→ main sequence (as a blue supergiant) → red giant → blue giant (only in some cases) → type II supernova→ neutron star. Less massive stars end up into white dwarfs, more massive ones into black holes.*

Oversimplifying the actual complex phenomenology, compact stars (white dwarfs, neutron stars and black holes) represent the final, generally stable stages of stellar evolution. Low- and intermediate-mass stars with $0.1 \lesssim M \lesssim 8\ M_\odot$ can only burn H into He and, if they are massive enough ($M \gtrsim 0.5\ M_\odot$), He into C and O. The inert core then contracts and, after expelling the extended H-He envelope that surrounds it (*planetary nebula)* and by this reducing the star's total mass below the Chandrasekhar's limit ($M_\text{Ch} \simeq 1.44\ M_\odot$), it settles down as a degenerate *white dwarf,* which slowly cools down and gets dimmer until visual extinction. Some white dwarfs, however, can accrete matter from a companion star, if they form a binary system; when the mass exceeds the Chandrasekhar's limit, the white dwarf collapses to eventually ignite runaway thermonuclear reactions in its center; the entire star is burnt to nuclear statistical equilibrium (mostly into iron) in a powerful explosion that completely disrupts the white dwarf, returning its matter to the interstellar medium (*type Ia supernova*). Very massive stars with $M \gtrsim 25\ M_\odot$ evolve rapidly through the main sequence and subsequent phases as hot, bloated and very luminous blue supergiants, often losing their extended envelopes already at early stages, due to strong stellar winds. At the end of their lives, they develop inert degenerate iron cores which are more massive than the mass limit for neutron stars (see Section V), so that their gravitational collapse cannot be halted by the degeneracy pressure of neutrons; either directly or after a supernova explosion, they eventually collapse into *black holes.* Predicted by General Relativity, these super-compact astrophysical objects represent the ultimate collapse of matter under the effect of gravity: the only properties left over of the original star are its total mass and angular momentum, and black holes can only be detected by the effects of their intense gravitational fields on the surrounding matter.

The stars which are expected to form neutron stars have masses in the range $8 \lesssim M \lesssim 25\ M_\odot$; they are massive enough that they can proceed through all stages of nuclear burning, all the way to Fe, but not too massive, so that the final neutron star is below its Chandrasekhar's limit and thence gravitationally stable. Since iron is the most stable nucleus, no further energy can be obtained from thermonuclear fusion; the inert iron core, embedded in the center of the extended and massive stellar envelope, starts contracting and becomes degenerate, the zero-point pressure of electrons holding it up against gravity. When its mass reaches the value $M_\text{Ch} \simeq 1.4\ M_\odot$ (Fe is continuously added to the core by shell-burning at its border), however, it collapses. The ensuing phenomenon*,* among the most energetic in the present Universe, is a *type II supernova*; they are also called core-collapse supernovae, to stress the physical mechanism that powers them: gravitation, not nuclear fusion [5].

Schematically, the core collapses, reaches at its center super-nuclear densities and consequently rebounds, due to the strong repulsive nature of nuclear forces at very small distances; incidentally, the physics of core collapse and rebound is also a fascinating subject for nuclear physics, since it involves *hot* and dense matter. The rebounding core creates a shock wave, that violently disperses into the interstellar medium the whole star envelope in an impressive electromagnetic display, the observed supernova explosion. The core itself, already very dense, hot and neutronized (proto-neutron star), contracts into its final neutron star state through the emission of a powerful burst of neutrinos, which in the case of SN1987A have been actually directly observed, thus confirming the general physical scenario. The newborn neutron star can remain in the system (like in the case of the Crab and Vela pulsars, observed in the middle of the corresponding supernova remnants), but it can also be kicked out, if the explosion is not spherically symmetric (which may be the case for SN1987A, where no pulsar has yet been detected more than 20 years after the explosion).

Notice how the core-collapse mechanism accounts for the presence of fast rotation and strong magnetic fields in young pulsars: during collapse, conservation of angular momentum and magnetic flux can increase the original $\omega$ and $B$ of the progenitor star by factors of $\sim 10^{10}$. The gravitational binding energy of the neutron star, $\sim GM^2/R \approx 10^{53}\ erg$, is mostly released in the neutrino burst: type II supernovae are actually neutrino bombs, the electromagnetic display representing only a tiny fraction of the energy available from the gravitational collapse.

## IV – HOW EXOTIC IS THEIR NUCLEAR STRUCTURE?

Having understood the origin and nature of neutron stars, we can now turn to describe in more detail the state of matter inside them. The underlying principle has already been explained in the previous section, namely the neutronization of matter with density. In the spirit of the present lecture, we only outline the main features and rationale of neutron star structure, referring the interested reader to the immense existing literature (the theoretical and experimental study of matter under extreme conditions is nowadays one of the most fascinating branches of nuclear physics) [6].

The radial profile of neutron stars is determined by hydrostatic equilibrium under gravitational forces and by the equation of state of matter as a function of density; since matter is strongly degenerate in most of the star, temperature can be neglected at first, as it will have little or no effect on the structure of a stable neutron star; temperature, however, is crucial when studying aspects like neutrino emissivity, transport of energy, cooling and so on. As a first approximation, we can take the Fermi model to represent the EOS of dense baryonic matter, namely neutrons are described as a gas of completely degenerate non-interacting fermions (like in the previous section). Even with such a simple EOS, the global features that characterize neutron stars appear clearly: for solar mass objects, Newtonian hydrostatic equilibrium yields densities ranging from zero at the surface to a few times $\rho_0$ at the center, over a radius of order $\sim 10^6 cm$. The Chandrasekhar's limiting mass, however, is quite large: $M_{\text{Ch}} = 5.76\, M_\odot$ (for pure neutron matter, the fraction of neutrons over nucleons is $Y_n = 1$), indicating the limits of such an approximation. Actually, two physical ingredients are necessary to obtain a realistic description of the star: first, an accurate EOS for matter at the different densities, which accounts for electrostatic interactions in the outer layers and for nuclear (weak and strong) interactions deeper in the star; then, General Relativity must be used to calculate the equilibrium (TOV equation), the gravitational fields being so strong that general relativistic corrections can no longer be ignored. The details of the resulting structure depend both on the mass of the star and on the EOS used to describe its matter; while the former is to be expected (more massive objects correspond to larger central densities, since they require more pressure to be stable), the latter is due to our still limited understanding of matter under extreme conditions (we will come back to this in Section V). However, a general outline of the radial structure of neutron stars can be given in terms of the increasing density.

Neutron stars are commonly divided into different regions (see Figure 5): the atmosphere and the surface, where the observed thermal spectrum is formed; the crust (sub-saturation densities), divided into *outer crust* and *inner crust,* of about $\sim 1\, km$ thickness and containing only a few percent of the total mass; the core (super-saturation densities), divided into *outer core* and *inner core,* of $\sim 10\, km$ thickness and including most of the star mass.

The surface (also called skin or envelope) is the region with $\rho < \rho_\beta$, namely below the critical value $\rho_\beta \simeq 10^7\, g\, cm^{-3}$ for neutronization; therefore it consists of normal matter in its ground state: $^{56}$Fe nuclei, crystallized in a Coulomb lattice. Its thermal, mechanical and electrical properties are reasonably well known, as it is the case for white dwarfs (the density range is the same, but the composition is different, since white dwarf matter has not reached nuclear equilibrium); they represent a fascinating study in solid state physics, due to the extremely large values of density, temperature and magnetic fields as compared to laboratory conditions.

Although less exotic and much smaller than the core, the crust is crucial for our understanding of the physics of neutron stars. On the one hand, it represents the interface between the observable surface and the hidden core: any physical information coming from the exotic matter in the center must go through and is affected by the crust before being observed. On the other hand, some peculiar observed phenomena (for example, pulsar glitches, thermal relaxation after matter accretion, quasi periodic oscillations, anisotropic surface cooling) may originate in the crust itself; we will come back to this in Section V. The outer crust is defined as the region $\rho_\beta < \rho < \rho_{\text{drip}}$ and the inner crust as the region $\rho_{\text{drip}} < \rho < \rho_{\text{core}}$, with $\rho_{\text{drip}} \simeq 4 \times 10^{11}\ g\ cm^{-3}$ the neutron-drip density and $\rho_{\text{core}} \sim 0.7\ \rho_0$ the density where "nuclei" (actually, at these densities, nucleon-cluster structures with exotic shapes, the so-called "pasta" phases) disappear. The equilibrium composition of matter both below and above neutron-drip has been extensively studied; one of the main features is the coexistence of neutron-rich exotic nuclei, which form a bcc (body-centered cubic) Coulomb lattice, in $\beta$-equilibrium with a strongly degenerate gas of relativistic electrons. As explained in the previous section, increasing density induces electron capture, but this takes place on nuclei rather than on free protons and thence nuclear force play a significant role in determining the equilibrium configuration: the nuclear species found in the outer crust have $Y_e = Z/A$ decreasing from $\sim 0.45$ ($^{62}$Ni) to $\sim 0.31$ ($^{118}$Kr). The other main features is the appearance of a sea of unbound dripped neutrons for densities above $\rho_{\text{drip}}$; the nuclei are so neutronized that a further increase in density causes neutrons to drip out of them. Thence, in the inner crust a new phase coexists in $\beta$-equilibrium with the Coulomb lattice of nuclei (at this point better called nuclear clusters, being immersed in the neutron fluid) and the degenerate electrons. This fluid of unbound neutrons becomes increasingly important with density; at $\rho \sim 10^{12}\ g\ cm^{-3}$ it contributes about 20% to the total pressure and the electron fraction is $Y_e \sim 0.15$, while at $\rho \sim 10^{13}\ g\ cm^{-3}$ it provides already about 80% of the pressure and $Y_e \sim 0.04$. Moreover, at these densities neutrons are expected to be paired by the nuclear residual interaction (S-wave Cooper pairs) and thence to be superfluid. We are in the presence of a unique scenario: a bulk, extended neutron superfluid, contained by gravity in a rotating vessel (the normal matter of the rapidly spinning star, namely nuclei and electrons coupled by the strong magnetic field); we will come back to the possible observational signature of this in Section V.

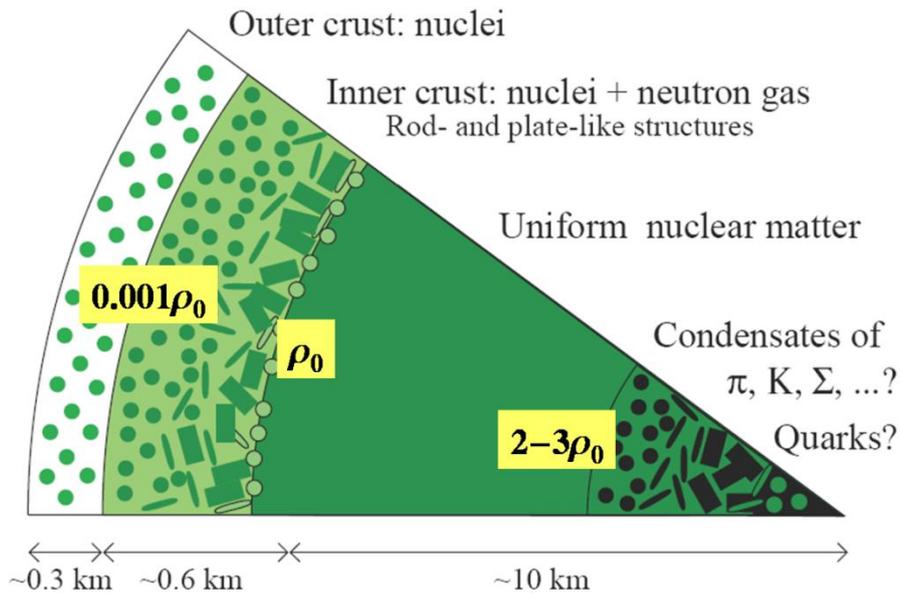

*Figure 5 – Schematic profile of a neutron star. The matter composition of the different regions, their approximate spatial extensions and the transition densities between the regions are indicated (from Ref. [6-a])*

The core of the star is the region with $\rho > \rho_{\text{core}}$, and here we are in the presence of bulk neutron matter, the clusters having melted at this point and neutronization being almost complete ($Y_e \sim 10^{-2}$). The neutrons are expected to form a P-wave superfluid and the few protons a S-wave superconductor, the latter strongly coupled to the magnetic field; again a unique scenario of bulk (kilometer-sized) quantum fluids made of paired nucleons. At $\rho \gtrsim 2\rho_0$, muons start to appear, since at these densities the electrons have Fermi energies of the order of the muon mass: the negative charge is now carried by degenerate electrons *and* muons, in $\beta$-equilibrium with the nucleons. The exotic scenario we have just outlined describes the outer core, which will be found in all neutron stars. For low-mass neutron stars, the outer core could actually constitute the entire core, since gravity is found to compress matter in their centers by less than twice saturation density. For more massive stars, however, most reasonable choices of EOS show that the density in their central regions could easily reach values larger than $2\rho_0$, up to several times saturation density. The critical densities for the appearance of new exotic phases of matter are model-dependent, so that the edge of the inner core cannot be uniquely defined. In general, we can say that for central density larger than $\rho_{\text{crit}} \sim 2 - 3\, \rho_0$ an inner core appears, which can present different exotic phases, produced by the very large Fermi energies of the particles in the strongly compressed matter. Among the possibilities, hyperonic matter (appearance of heavy baryons), meson condensates (pions or kaons), *deconfined* degenerate quark matter (possibly in a state of color-superconductivity). The last one is particularly exciting, since it makes neutron stars also relevant for QCD: actually, the phase space of hadronic matter in the low-$T$, large-$\rho$ region can only be tested by the detection and identification of these (still hypothetical) compact objects, either entirely or only centrally composed of deconfined quarks (respectively *strange* and *hybrid* stars). We will come back to this in the next section.

**V – WHICH OBSERVATIONS COULD BE RELEVANT TO NUCLEAR PHYSICS?**

At this point, it is clear that neutron stars represent unique objects from any directions we look at them. The question is, can we probe and study in more detail their exotic interior by astronomical observations of their external properties? Although difficult, this is actually possible and represents an exciting and active field of research in astronuclear physics. Several and at times unexpected connections have been found between macroscopic astrophysical properties, that are already or can eventually be detected, and the microphysical state of matter hidden inside these stars. The list is long and probably useless at the introductory level of this lecture; the potential observational signatures range from mechanical to thermal to electromagnetic features, and involve sophisticated astronomical observations in all bands, as well as future detection of neutrinos and gravitational waves. Here we will just outline three significant examples, that provide a glimpse of the fascinating interplay between large- and small-scale properties of neutron stars [6].

**V.I - Mass, radius and EOS**

In Section III, we discussed the Chandrasekhar's mass; the value $M_{\text{Ch}} = 5.76\, Y_e^2\, M_\odot$ (obtained for non-interacting degenerate matter in equilibrium under Newtonian gravity) is quite reasonable for white dwarfs. Indeed, the electrostatic corrections due to the interaction between electrons and ions turn out to be quite small in these objects, and general relativistic effects are altogether negligible. The situation for neutron stars, however, is completely different: in this high-density environment, nucleons interact strongly through the nuclear forces; moreover, the Newtonian limit is not valid anymore in these very compact objects, and General Relativity must be used to study gravitational stability and determine the maximum allowed mass. As it turns out, the Chandrasekhar's mass for neutron stars depends significantly on the EOS used to describe baryonic matter, namely its value is determined by the effect of nuclear forces. In this sense, therefore, mass measurements can yield information about and give constraints over the EOS of super-dense matter.

In Figure 6, we show a mass-radius diagram for neutron stars. Some regions are forbidden for gravitational stability, and this can be due to General Relativity (the star would collapse into a black hole), causality (the EOS would allow sound waves with velocity faster than light) or rotational limits (the star would fly apart, gravity not being enough to provide the centripetal acceleration). The curves labelled with letters represent the mass-radius relation obtained for different EOS: the curves to the right of the figure are for normal nuclear matter (neutron stars), those to the left for quark matter (strange stars). Neutron stars (and hybrid stars) are bound by gravity, and their radius decreases for increasing mass; strange stars are bound by gluons, and their radius increases with mass. For the same mass, strange stars are more compact than neutron stars and thus they are able to sustain faster rotations.

The potential of the mass-radius diagram is easily explained. For example, a one solar mass pulsar with a period of half a millisecond would unequivocally identify it as a strange star; in order to satisfy the Keplerian limit ($\omega < \omega_k = \sqrt{GM/R^3}$), such an object should have a radius of less than ten kilometers: from Figure 6, we see that this is only consistent with a quark matter EOS. Similarly, unequivocal observation of a neutron star with $M > 2\,M_\odot$ would rule out most of the soft EOS, leaving as possible only the stiffer ones (in the present case, those labelled MS0 and AP4 [3-a]), which are compatible with such large masses. The best case would be the direct measurement of mass *and* radius for the same star; this would not only constrain the possible classes of EOS, but it would precisely indicate a point in the diagram where the theoretical mass-radius curve must pass, thus picking out a particular EOS out of the several which have been constructed, using different prescriptions for the nuclear forces or different approximations to describe the extended system of strongly interacting nucleons (or hadrons in general). Although so far it has not been possible to determine mass *and* radius (the latter in particular) for the same neutron star with enough precision, the possibility in the future of such an observation would be invaluable to nuclear physics, providing a big step forward in our understanding of the nature of nuclear forces.

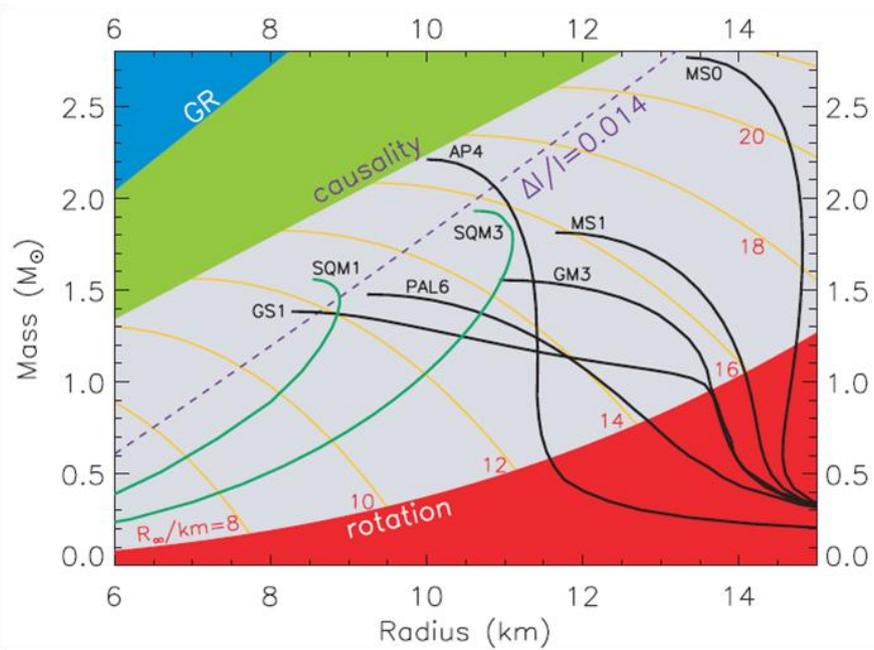

*Figure 6 – Mass-radius relation for different EOS of dense matter. The curves to the right of the figure are for normal matter, those to the left (labelled SQM) for quark matter. Regions of the diagram excluded by rotation, causality and General Relativity are also given (from Ref. [3-a]).*

**V.II – Cooling of neutron stars**

The launch of more powerful x-ray satellites in the last decade has made possible the study of the surface temperature of several isolated neutron stars. When coupled with an estimate of the age of the star, these observations provide an alternative way to investigate the properties of the exotic interior. Very dense and/or very hot matter cools mainly by neutrino emission (lepton cooling), rather than through the transport of heat by photons or electrons (radiative and conductive cooling). Neutrino cooling is crucial in the last stages of evolution of massive stars, in the gravitational collapse of their inert iron core, in the evolution of the proto-neutron star as it contracts into its stable compact configuration, and in the cooling of the young neutron star. Neutrino emission is an extremely efficient cooling mechanism, which draws energy directly from the dense and hot core; neutrinos emitted in the centre of a neutron star escape freely out into space without interacting with matter, as opposed to the more common radiative or conductive processes, where heat carried by photons or electrons diffuse slowly from the centre to the surface, transferring energy which will eventually be radiated into space.

The result is that the crust of a young neutron star, where slow conductive cooling dominates (heat transfer is mostly conductive when degenerate electrons are present), is soon much hotter than the core, which cools by lepton emission. The subsequent thermalization of the star, where heat diffuses *inward* from the crust to the core, determines the cooling history of the observable radiating surface. Seen from another angle, we can say that the thermal signature from the cold interior must diffuse through the insulating hotter crust around it, before reaching the surface and being detectable by an outside observer. The purpose of cooling studies is to calculate the observed surface temperature and luminosity of neutron stars as they age, by detailed numerical simulations; cooling profiles obtained for different physical input parameters are then compared to observations, as shown in Figure 2b. Several important factors determine the shape of the cooling curves: mass, EOS, neutrino emissivities, specific heats, thermal conductivities, pairing properties; therefore, a systematic analysis can be carried on, to confirm or rule out some of the many possible scenarios and thus test indirectly the state of matter inside neutron stars. Next, we give two examples of how cooling is affected by the microscopic state of matter and can thus be used as a diagnostic tool.

In Figure 7a, we show two typical theoretical cooling curves: the surface temperature (left scale) and luminosity (right scale), as observed by a distant observer (at infinity), are given as a function of the age of the neutron star. Three temporal regions can be individuated, which characterize different cooling phases: the core relaxation epoch (I), the neutrino cooling epoch (II), and the photon cooling epoch (III). During core relaxation, the surface temperature is practically constant; although the core has already cooled by neutrino emission and become isothermal, the thermal inertia of the surrounding crust still prevents the thermal signal to reach the surface and affect its temperature; the length of the plateau phase (in the range $10^1 - 10^2 \, yr$) depends on the thermal and pairing properties of the crust. During epoch II, the surface responds to the cooling of the deep interior and its temperature decreases; the overall loss of energy from the star is still dominated by neutrino emission from its denser parts. After $\sim 10^5 \, yr$, the interior of the star has cooled enough that neutrino emission becomes inefficient: during epoch III, the cooling of the star is determined by emission of photons from the surface, with standard conductive heat transfer inside; the different cooling mechanism explains the different slope of the curves in regions II and III.

The two curves in Figure 7a refer to two different types of neutrino cooling in the stellar core: *standard cooling* processes are slower and take place in standard neutron matter, namely in the outer core; *rapid cooling* processes are very fast and they are associated to the different exotic states of matter (meson condensates, hyperonic matter or quark matter), which appear if the central density is large enough for the formation on an inner core. Altogether, standard cooling would indicate low-mass stars with a stiff EOS, while rapid cooling would be typical of either low-mass stars with a soft EOS or high-mass neutron stars. As a product of the supernova explosion, the core is formed at $T \sim 10^{11} - 10^{12} \, K$; it is so hot that matter is only weakly degenerate and it cools to temperatures $T \sim 10^9 \, K$ in a few days, by intense neutrino emission from its non-degenerate matter (the so-called direct-URCA process). When $T \lesssim 10^9 \, K$, matter in the core becomes degenerate and the rate of neutrino cooling is determined by the composition of matter, which in turn depends on the density reached under gravitational compression. For normal neutron matter, only standard cooling (the so called modified-URCA process) is allowed; in this regime, the core cools to $T \sim 10^8 \, K$ in a few centuries. For exotic matter, instead, fast cooling (analogous to direct-URCA) is allowed and the core can reach temperatures as low as $T \sim 10^7 \, K$ in just a few decades.

The observational signature for the existence of an inner core of exotic matter is evident from Figure 7a: a sudden drop in surface temperature of a young neutron star would be unequivocal evidence of the presence of such a core; moreover, older stars which are too cold, as compared to the standard cooling results (e.g., Vela in Figure 2b), could also indicate the presence of exotic matter in their interiors. Incidentally, the time of rapid cooling, $t_w$, if ever observed could also constrain the EOS of dense matter [7].

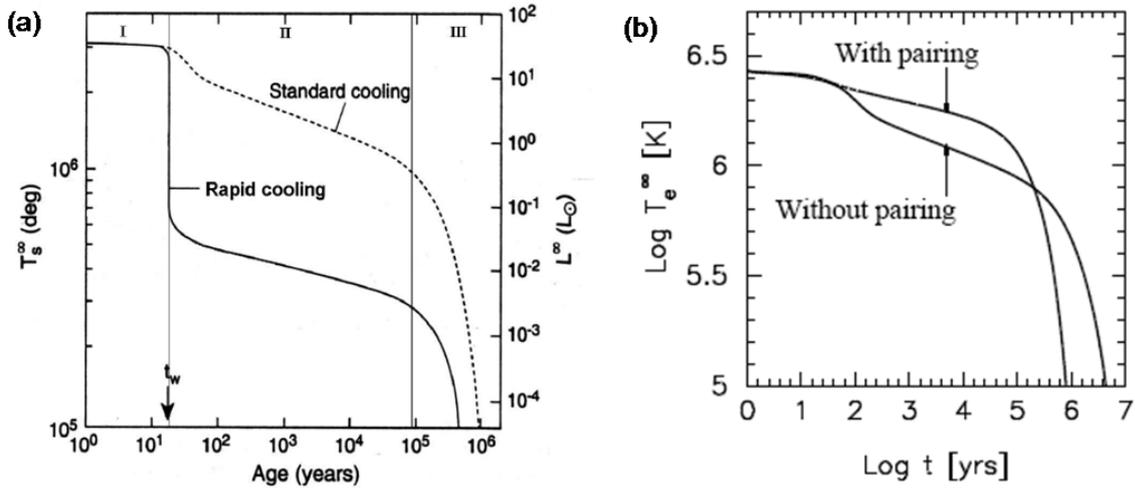

*Figure 7 – (a) Typical cooling curves for standard (dashed) and rapid (solid) cooling of the core. The three vertical regions correspond to:* I) *core relaxation epoch;* II) *neutrino cooling epoch;* III) *photon cooling epoch (from Ref. [7]). (b) Example of standard cooling curves calculated with and without nucleon superfluidity in both the core and the inner crust.*

As a second example, we consider Figure 7b; here, the two cooling curves are obtained from standard cooling calculations *with* or *without* nucleon superfluidity in the core and inner crust. Pairing affects cooling in two ways: it quenches neutrino emissivities and modifies the specific heat of superfluid nucleons. Although the figure corresponds to a particular choice of pairing parameters (among the several combinations possible) the general effect is evident: the thermal evolution of the surface is seen to be quite sensitive to the superfluid properties of matter, both in the core and in the inner crust, and cooling can be used to probe indirectly such properties.

**V.III - Pulsar glitches**

Pulsars show a regular slow down in their rotation, due to the emission of energy at the expenses of the rotational kinetic energy. Several pulsars, however, present sudden spin ups called *glitches*; the Vela pulsar, for example, has been glitching once every 2-4 years for the last thirty years, with jumps of the order of $\Delta\omega/\omega \sim 10^{-6}$ and $\Delta\dot{\omega}/\dot{\omega} \sim 10^{-2}$ (see Figure 8a). To explain this phenomenon, which involves a massive and sudden transfer of angular momentum to the emitting crust, several models have been proposed, with different internal mechanisms to store for a few years and then rapidly release the angular momentum needed to cause a glitch.

One of the most promising and intriguing explanation of pulsar glitches is the so-called *vortex model*, which is based on the coexistence of a neutron superfluid and a lattice of nuclei in the inner crust of rotating neutron stars. The mechanism invokes an important characteristic of rotating superfluids, which is well known and observed in the laboratory for $^4$He and $^3$He: a superfluid cannot rotate as a rigid body, since superfluid flow must be irrotational, while rigid rotations with $\vec{v} = \vec{\omega} \times \vec{r}$ have $\vec{\nabla} \times \vec{v} = 2\vec{\omega} \neq 0$. Superfluids in a rotating vessel, instead, develop an array of quantized vortex lines which are parallel to the rotational axis of the container: these represent singularities in the rotational flow (a line vortex has a velocity profile $v(R) \propto 1/R$ where $R$ is the distance from the vortex axis), and they carry quantized angular momentum. Quantistically, vortices are excitations of macroscopic energy and angular momentum, which locally (in the vortex core) destroy superfluidity and thus allow a coupling of the superfluid to normal matter (e.g., electrons can scatter off the vortex cores). The dynamics of the superfluid must then be described in terms of distribution and motion of vortices, and an *average* superfluid velocity can be defined in terms of the number of vortices. The superfluid can decrease its angular momentum only by expelling vortex lines from its interior; if for some reason (e.g., impurities) the lines cannot move, the angular momentum and rotational pattern of the superfluid are *frozen* and its average velocity cannot change.

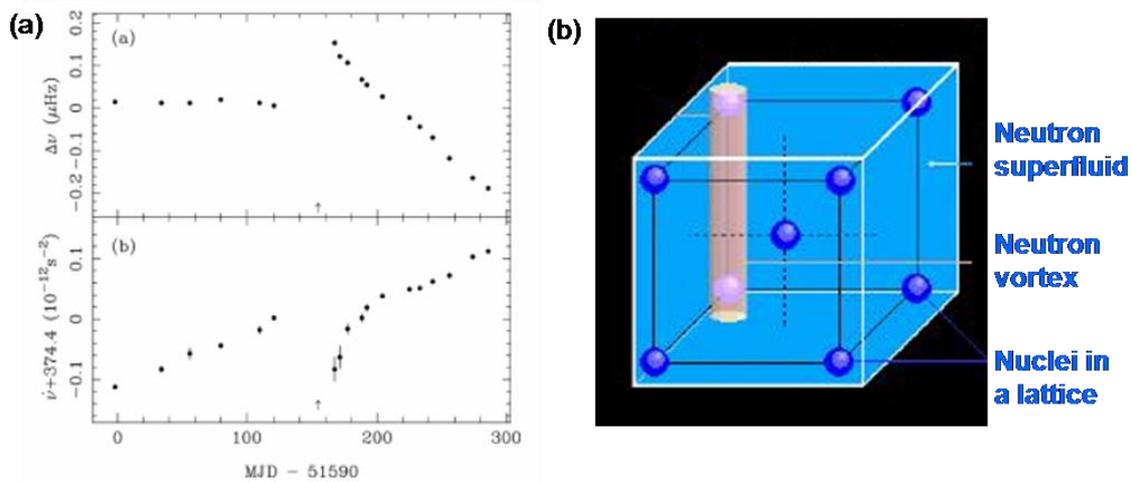

*Figure 8 – (a) Observations of a pulsar glitch: the sudden jump is detected both in the rotational frequency and in its time derivative. (b) Interaction between a vortex in the neutron superfluid and the lattice of exotic nuclei in the inner crust of neutron stars.*

As discussed in Section IV and shown in Figure 8b, the inner crust of neutron stars is characterized by a superfluid of unbound neutrons and a Coulomb lattice of exotic nuclei (clusters). The normal matter of the star (nuclei and electrons, coupled by the strong magnetic field throughout the star) forms a rotating container, so that the neutron superfluid in the inner crust develops an array of vortices, whose number density is determined by the angular velocity of the star (Feynman-Onsager relation [8]). The nuclear lattice, instead, represents a system of impurities (the nuclear clusters) to which the vortices can potentially *pin* and thus lose their mobility. In the case of complete pinning of the vortex lines to the lattice, the following situation takes place: the crust and the vortices pinned to it slow down as the star looses rotational energy, but the neutron superfluid is frozen and its average velocity cannot follow the slow down. An increasing velocity lag between the superfluid and its vortices then develops, and in turn this applies strong lift forces (Magnus force) on the vortex lines, which tend to unpin them from the lattice of nuclei. When the Magnus force, which is proportional to the velocity lag and thence increases with time, becomes larger than the pinning force which binds the vortex lines to the lattice, the vortices catastrophically unpin from the lattice and transfer their angular momentum to the surface, thus spinning up the star and causing a glitch.

We do not even try to go into any detail, the model being quite complex and requiring specific knowledge of vortex dynamics and interactions; although very promising, it has not yet proven to be the correct explanation, also because some fundamental questions are still open. For example, neutron vortices also form in the core of the star, and here they interact with the magnetic flux tubes associated with proton superconductivity: this is still a quite unchartered issue. Even more crucial, the pinning force that binds the vortex lines to the lattice of exotic nuclei is still poorly known: the interaction of a vortex with a *single* nucleus in the neutron-drip regime is already a hard problem to treat quantistically, and extending the calculation to the whole lattice is even more forbidding. However, reasonable approximations are also possible, and the results so far are not inconsistent with observations. The main point we want to stress here is the following: if the vortex model is the correct explanation, pulsar glitches would be unique and direct macroscopic *evidence* of the existence inside neutron stars of an extended (kilometre-sized) neutron superfluid, which interacts significantly with the nuclear lattice via its vortex lines. The idea is fascinating and allows an independent way to probe some nuclear properties of matter (e.g., neutron pairing and structure of exotic nuclei beyond neutron-drip) in the inner crust of neutron stars.

**VI – CONCLUSION**

We have reached the end of this lecture. Although only a first introduction to the physics of neutron stars, it shows how the main properties of these compact objects follow directly from general physical principles: when matter is compressed by gravity to its extreme limits (i.e., short of becoming a black hole), which can happen only as the result of the long evolutionary process of massive stars, the final object naturally shows extreme physical conditions. Better understanding of neutron stars can come from an interplay of observations, astrophysics and nuclear physics. Since observations of macroscopic features of the star must be interpreted in terms of the microphysics in their interiors, a strong interplay of theorists from different areas of physics is necessary: the European network *Compstar* is an example of such an interdisciplinary approach [2]. From the point of view of observations, the beginning of the new millennium is a particularly exciting time; satellite-based observatories in different bands are giving our scientific community an incredible wealth of new data: several new classes of neutron stars have been discovered in the past decade, and the first results from the recently launched *Fermi* $\gamma$-telescope appear even more promising. Moreover, the now fully operational detectors of gravitational waves (e.g., *Virgo* in Italy and *Ligo* in the US) could soon open a new window for the investigation of compact objects (astro-seismology by gravitational waves), while large neutrino detectors (e.g., *SuperKamiokande*) make the possibility of extra-solar neutrino astronomy (started with SN1987A) a concrete option. Finally, heavy-ion colliders (e.g., *RHIC* and now *LHC)* will enable nuclear scientists to probe some properties of exotic nuclei; although the conditions found inside neutron stars (even those in the inner crust) are well beyond the reach of any experimental facility, direct data about the isospin-asymmetric matter formed in heavy-ion collisions could help nuclear theorists to understand whether they are on the right track. Only from a synergic effort of all these different fields of expertise we will be able to reach a coherent description of neutron stars: a unique environment in our present Universe, where gravitation dominates over the other fundamental interactions and, by compressing matter to its limits, determines extreme physical scales and exotic scenarios, both at the macroscopic and microscopic level.

**ACKNOWLEDGEMENTS**

This work was partly supported by CompStar, a Research Networking Programme of the European Science Foundation. Some of the figures were taken from the web. This lecture was given at the Ecole Joliot Curie 2009, in the nice atmosphere created by the organizers and participants of the school at Lacanau.